\documentclass[jr]{sjp} 
\usepackage{graphicx} 
\usepackage{amsmath,amssymb,subfigure}
\usepackage{epsfig}
\begin{document}
\title{Magnetic Skyrmions in Condensed Matter Physics}
\author[]{Sayantika Bhowal,$^\ast$\footnote{$^\ast$ bhowals@missouri.edu} \  S. Satpathy, \   and Pratik Sahu}
\address[]{Department of Physics \& Astronomy, University of Missouri, Columbia, MO 65211, USA \\}

\begin{abstract}
Skyrmions were originally introduced in Particle Physics as a possible mechanism to explain the stability of particles.
Lately the concept has been applied in Condensed Matter Physics to describe the newly discovered topologically protected 
magnetic configurations called the magnetic Skyrmions. This elementary review introduces the concept at a level suitable for beginning students of Physics.
\end{abstract}
 
\maketitle

\section{Introduction}
The basic objects of Classical Mechanics are stable particles, characterized by a non-zero mass, which live for ever. 
In contrast, at a fundamental level, elementary particles such as electrons and protons are described using quantum field theory, where they are thought of as wave-like excitations of an underlying field. 
%Particle Physics in contrast decribes the elementary particles such as electrons and protons to be wave-like excitations of an underlying field. 
It is however a non-trivial task in  field theory to make these wave-like excitations stable; they would generally dissipate similar to the disappearance of the waves in a pond once they have been  created. In the year 1962, Skyrme \cite{Skyrme} proposed the idea that the particles do not decay because they are topologically protected, in the sense that they have a topological number, which can not be changed by a continuous deformation of the underlying field. 
In topology, a donut is topologically equivalent to a cup, as they both contain one hole, and one can be continuously deformed to the other. A sphere contains no hole and is topologically different, and thus can not be deformed into a donut by continuous deformation.
Sometime back, in the year 1989, it was pointed out that topologically protected systems may be relevant in  condensed matter physics \cite{Bogdanov1989}.
Recently, there has been a flurry of theoretical and experimental works \cite{Nagaosa} showing that topologically protected states can be stabilized in chiral magnets in the form of a swirling spin texture called a ``magnetic Skyrmion."
The present review aims at introducing the basic concepts of the magnetic Skyrmion to beginning students of Physics.

%
%:Fig 1
\begin{figure}[h]
 \begin{center}
\includegraphics[scale=.4]{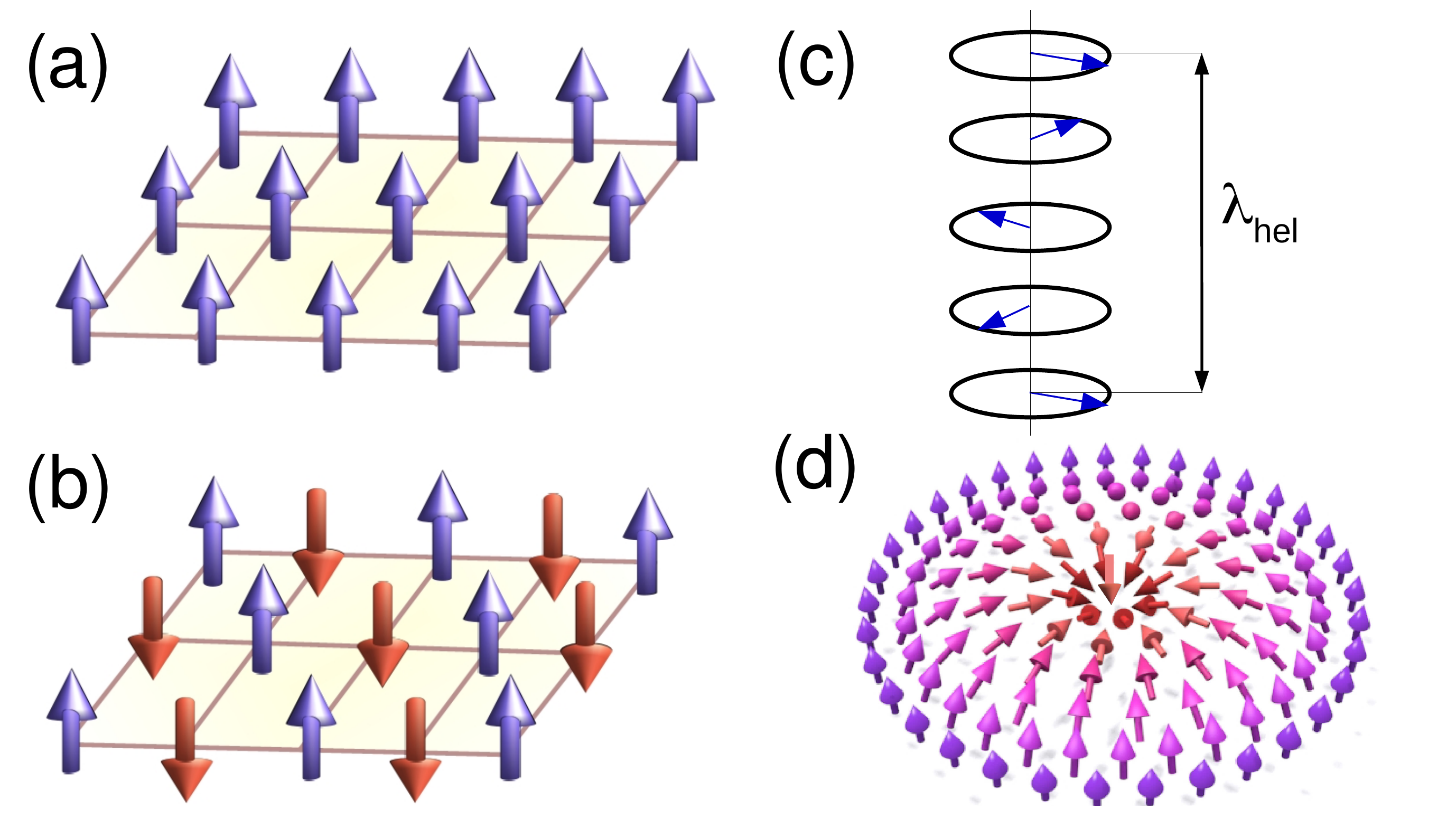}
\caption{Different types of magnetic states in solid. (a) Ferromagnetic state, (b) Anti-ferromagnetic state (c) Helical state, and (d) Skyrmion state.
For the ferromagnetic and the anti-ferromagnetic structures,  magnetic moments of individual atoms are arranged as shown. In the helical state, ferromagnetic planes are stacked on top of one another, with their magnetic moments turning with the periodicity
determined by the helical length $\lambda_{\rm hel}$. The Skyrmion state is characterized by a certain Skyrmion number as discussed in the text.
(Fig. (d) is reproduced with permission from the authors of Ref. \cite{Jacob}.)
}
\end{center}
\label{fig1}
\end{figure}

\section{Magnetic interaction in solids}
In magnetic solids, such as Iron, atoms acquire a magnetic moment due to the spin of the electron, or in some cases due to its 
orbital motion. Arranged on a periodic lattice structure in the solid, the atomic moments interact with one another, leading to well defined magnetic structures such as a ferromagnet or an anti-ferromagnet.
The simplest Hamiltonian that describes the interaction between the magnetic moments is 
 the Heisenberg Hamiltonian 
\begin{equation}\label{Heisenberg}
 {\cal H}_H= J \sum_{ij} \vec{S}_i\cdot \vec{S}_j,
\end{equation}
where $\vec{S}_i$ is the (spin) magnetic moment at site $i$ and $J$ is the interaction between nearest neighbors. 
The sign of $J$ determines the relative alignment of the spins, e.g., if $J < 0$,  the spins prefer to align parallel to each other (ferromagnet) so that the energy is a minimum. On the other hand, a positive $J$ leads to an antiparallel alignment resulting in an antiferromagnetic structure (See Fig. \ref{fig1}). In solids with strong spin-orbit coupling and broken inversion symmetry, a cross-product or chiral interaction, 
known as the Dzyaloshinskii-Moriya (DM) interaction exists. Although much smaller in magnitude than the 
Heisenberg interaction, the DM interaction can lead to fundamentally new Physics.
The DM interaction reads
\begin{equation}\label{DM}
 {\cal H}_{\rm DM}=\sum_{ij} \vec D_{ij} \cdot (\vec{S}_i\times \vec{S}_j),
\end{equation}
where $\vec D$ is the strength of the DM interaction.
In contrast to the Heisenberg interaction which only favors parallel or antiparallel alignment of the spins, 
the DM interaction favors a canted spin alignment (energy becomes minimum if the two spins are perpendicular to each other and
lie on a plane normal to $\vec D$, as can be immediately seen from Eq. \ref{DM}).  

%: Fig 2
\begin{figure}[t]
 \begin{center}
\includegraphics[scale=.30]{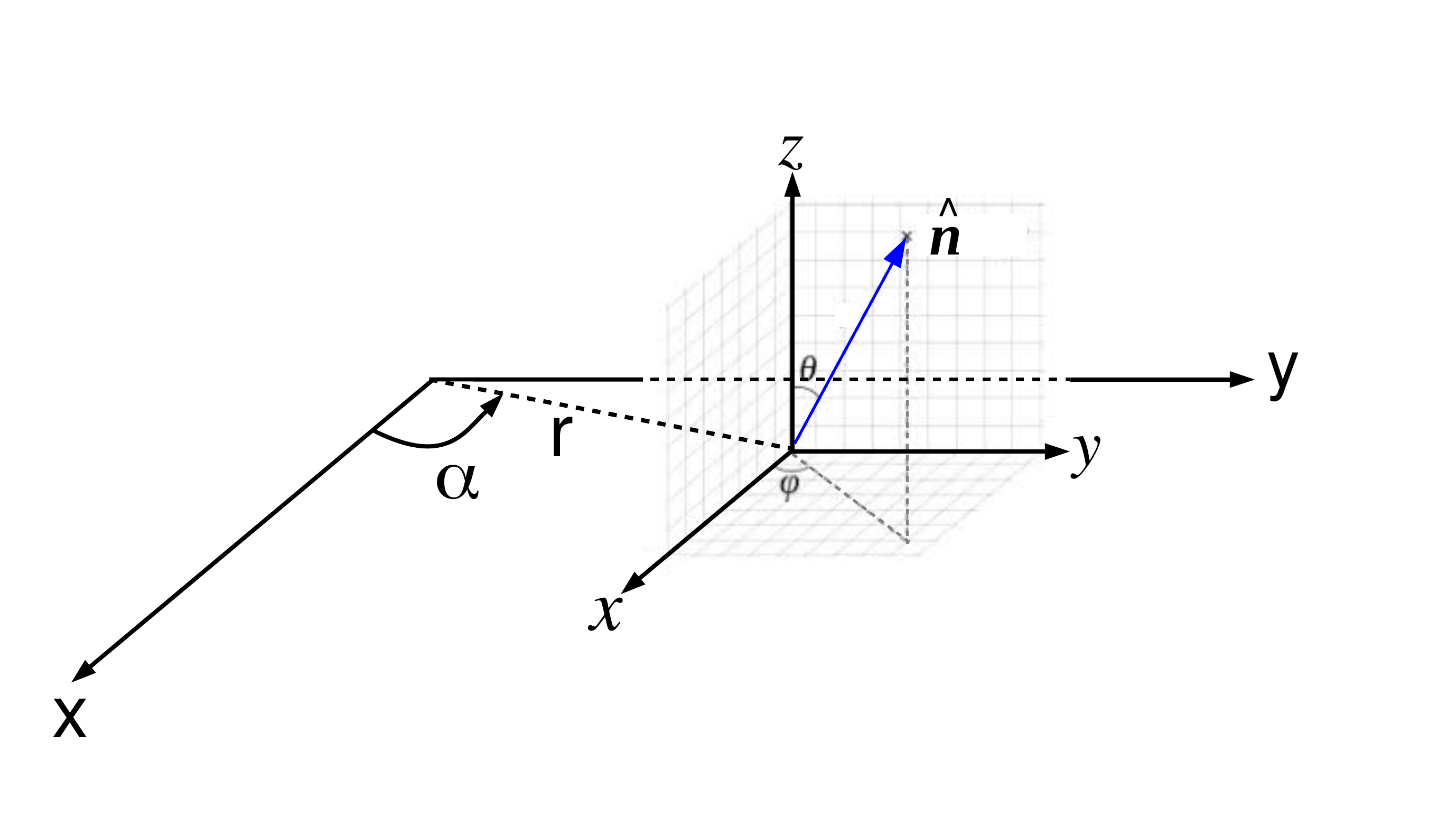}
\caption{Polar coordinates to describe the spin texture in the continuum model of the Skyrmion state.
The direction of the magnetization vector $\vec n (\vec r)$ depends on its location $\vec r$ on the 2D plane, described by the polar coordinates $(r, \alpha$), but $\vec n (r, \alpha)$ can point anywhere in 3D, 
described by  the spherical coordinates $\theta $ and $\phi $ at each point $\vec r$.
For the skrmion state, the polar angle $\theta (r)$ and the azimuthal angle $\phi (\alpha)$ are functions of only $r$ and $\alpha$, respectively.  
An electron moving in the spin texture $\vec n (\vec r)$ experiences a topological electric amd  magnetic field, caused by the fact that the electron spin must follow the local spin directions as it moves in the 2D space.
}
\end{center}
\label{polar}
\end{figure}

The presence of both the  Heisenberg and the DM interactions leads to a helical spin state, and the DM interaction under certain conditions can lead to the Skyrmion state as well, which we will discuss shortly. 
In the simplest helical state, ferromagnetic planes of spins turn along the direction normal to the planes, with a helicity angle $\vartheta$ and the helical period 
$\lambda^{\rm hel} = 2\pi/\vartheta$ (Fig. \ref{fig1} c). The helicity angle is determined by the minimization of the net energy, coming directly from the expressions for the Heisenberg and the DM interactions,
\begin{equation}\label{Helicity}
 E (\vartheta) = JS^2\cos \vartheta + DS^2 \sin \vartheta,
\end{equation}
which leads to  $\vartheta = \tan^{-1} (D/J)$. 
The helical period ($\lambda^{\rm hel}$) in some representative materials are listed in Table \ref{tab2},
which is indicative of the strength of the DM interaction.
%, a mechanism that is often responsible for the formation of the Skyrmion state.

%:Table 1
 \begin{table} [h]    
\caption{N\'eel temperatures (T$_N$) and helical periods ($\lambda^{\rm hel}$) in some of the magnetic materials with broken inversion symmetry.}
\centering
\setlength{\tabcolsep}{12pt}
 \begin{tabular}{c c c}
\hline
Materials & T$_N$ (K) & $\lambda^{\rm hel}$ (nm)\\
%            & ($\frac{\hbar}{e} \Omega^{-1}$cm$^{-1}$) \\
\hline\hline
MnSi (Bulk) & 30 & 18 \\
MnSi (Thin film) & 45 & 8.5 \\
MnGe (T = 20 K) & 170 & 3 \\
Mn$_{0.5}$Fe$_{0.5}$Ge & 185 & 14.5 \\
\hline
\end{tabular} 
\label{tab2}  
\end{table}

\section{ The Skyrmion state} 
The Skyrmion state is a complex magnetic state, the formation of which is mediated by one of several mechanisms \cite{Garel,Heinze,Okubo,Brey}, viz., (i) The DM interaction together with a magnetic field, (ii) Magnetic dipolar interaction, (iii) Frustrated exchange interaction, (iv) Four-spin interactions, and (v) Rashba spin-orbit coupling in the presence of itinerant electrons (polaronic Skyrmion). 

%which forms under certain situations when additional interactions are present or when an external magnetic field is present. Although we don't get into the details, several mechanisms have been proposed to mediate the formation of the Skyrmion state, 

%: Fig 3
\begin{figure}[t!]
\label{fig-expt}
 \begin{center}
\includegraphics[scale=.55]{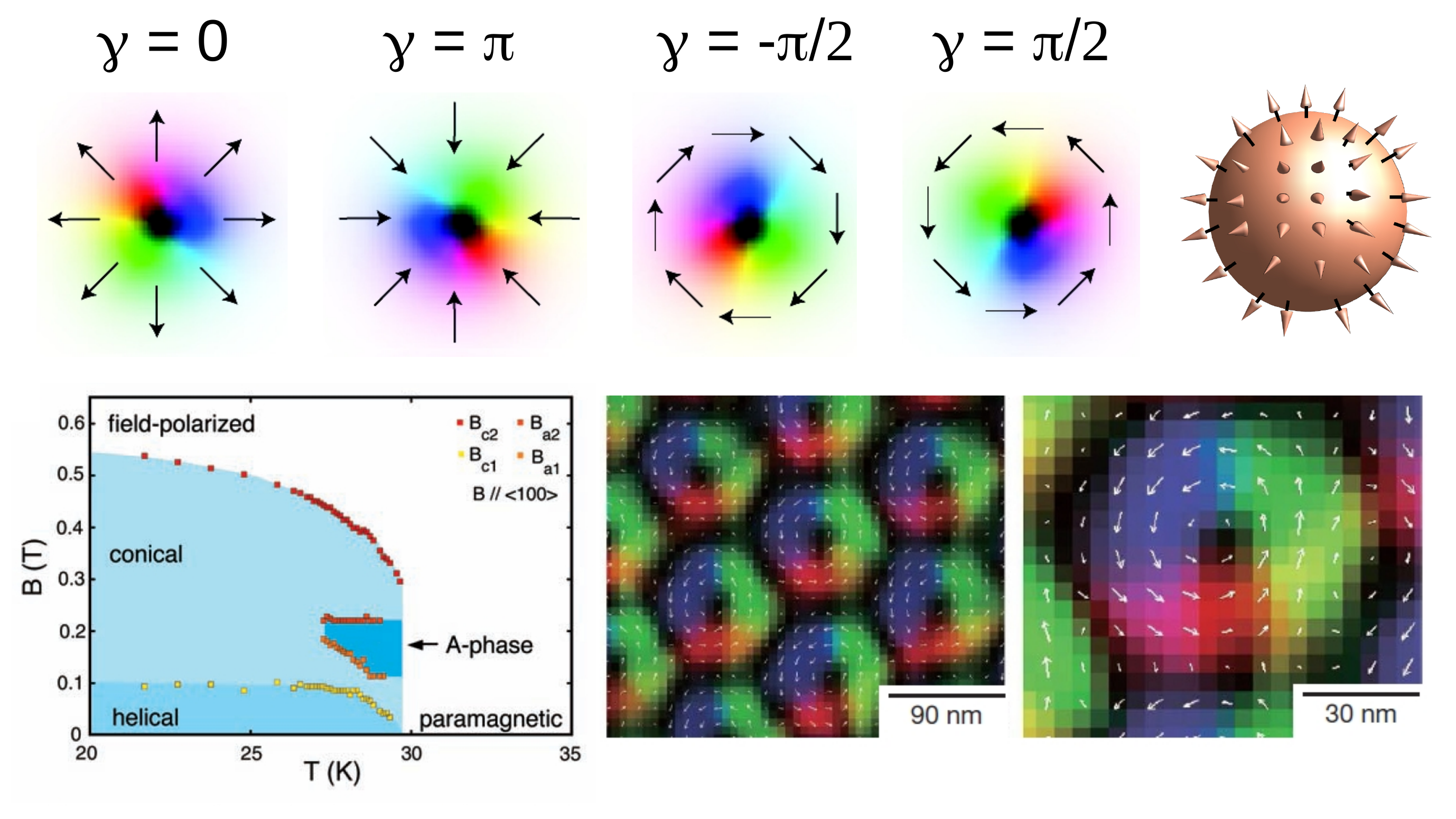}
\caption{{\it Top panel:} Several Skyrmion spin structures corresponding to the winding number $m = 1$ and helicity parameter $\gamma = 0, \pi, \pm \pi/2$. 
The in-plane components of the spins are indicated by arrows,
 while the color map denotes the out of-plane component of the spin, where the black and white colors represent the down and up spin state, respectively. 
{\it Top panel right:} 
 The spherical ``hedgehog" of spins, the stereographic projection of which leads to the 2D
magnetic Skyrmion as discussed in the text. The figure corresponds to the winding number $m = 1$.
{\it Bottom  left:}  The phase diagram of MnSi in the parameter space of temperature (T) and magnetic field (B), showing the stabilization of the Skyrmion phase (the so called ``A-phase") with small external magnetic field. {\it Bottom middle:} The real space configuration of the Skyrmion crystal in Fe$_{0.5}$Co$_{0.5}$Si subjected to a weak magnetic field. {\it Bottom right:} The magnified view of this Skyrmion spin texture. The direction of the spin at each point is described by the color map: black color denotes $\uparrow$ or $\downarrow$ spin and the white
arrows  denote the in-plane component. Figure reproduced with permission from: top panel, Ref. \cite{Nagaosa} \textcopyright~2013 Springer Nature; bottom left, Ref. \cite{MnSi}   \textcopyright~2009 AAAS;   bottom middel and right, Ref. \cite{Yu2010}  \textcopyright~2010 Springer Nature.
}
\end{center}
\end{figure}
 In addition, the Skyrmion state is defined such that $\theta (r)$ changes from $\theta = \pi$ at the center to $\theta =0$ at the boundary of the Skyrmion ($r =\lambda$) or vice versa. 
 Single-valuedness of the spin orientation demands that the azimuthal angle is of the form 
 \begin{equation} 
 \phi = m \alpha + \gamma,   \hspace{10mm}   ( {\rm winding \  number \ {\it m} \  }
 {\rm and \  helicity \   parameter \ }   \gamma
 {\rm \ defined)}
 \end{equation}
A magnetic Skyrmion is a swirling magnetic structure of spins as illustrated in Fig \ref{fig1}.  
It is usually a two-dimensional (2D) object, existing at interfaces between two materials or in magnetic thin films.
The topological properties depend on the geometry of the structure. 
The two key characteristic geometrical quantities are the vorticity and the helicity
of the structure, which are a characteristic of how the spin orientation $\vec n(\vec r)$ changes over space. 
Note that for the lattice, the position $\vec r$ takes discrete values, while in continuum models it is a continuous variable.
The spin orientation  at each point $\vec r \equiv (r, \alpha)$ on the 2D plane is described in the spherical coordinates 
by specifying the polar and azimuthal angles, $\theta (\vec r)$ and $\phi (\vec r)$, respectively, as indicated in Fig. \ref{polar}. 
For the Skyrmion state, 
$\theta (\vec r)$ is a function of the radial distance $r$ only and $\phi (\vec r)$ is a function of the polar angle $\alpha$ only, so that in the cartesian coordinates, the local magnetization vector is
\begin{equation} \label{n}
 \vec n(\vec r) = (  \sin\theta(r)  \cos\phi(\alpha),  \sin\theta(r)  \sin\phi(\alpha),\cos\theta(r)). 
\end{equation}
 where $m$ is  a non-zero integer called the winding number. 
 Skyrmions and anti-Skyrmions are defined as those for which the winding number is positive or negative, respectively. %\cite{Blugel2017}.
 The helicity
 parameter $\gamma$ takes specific values for helical states. If $\gamma = \pm \pi/2$, then the helicity $h = \pm 1$ 
 (the two signs indicate left or right handedness), while for $\gamma = 0$ or $\pi$, we have a radial spin structure. The Skyrmion spin texture for the winding number $m = 1$ and for different helicity parameters $\gamma = 0, \pi, \pm \pi/2$ are shown in Fig. \ref{fig-expt} 
 for illustration.

Physically speaking, a 2D magnetic Skyrmion can be thought of as a topological object which is formed
by a stereographic projection from a spherical ``hedgehog" of spins as shown in Fig. \ref{fig-expt},
where the down spin at the south-pole is mapped onto the center of the 2D disk, while the up spin at the north-pole is mapped onto the edge circle of the disk, far off from the centre. 
The Skyrmion number $N_{sk}$ denotes the number of times $\vec n (\vec r)$ wraps around the unit sphere. 
As shown in Section \ref{sec6}, the Skyrmion number simply equals the winding number $m$ apart from a sign.
Note that for trivial magnetic structures, e.g., ferro and antiferromagnetic order, $N_{sk} = 0$, while for the example shown in Fig. \ref{fig-expt}, the Skyrmion number is one. 

{\it Experimental observation} -- 
Skyrmions were first observed in the year 2009 in  MnSi using neutron scattering
 \cite{MnSi},
when the well-known A-phase of MnSi was identified as the  Skyrmion phase (Fig. \ref{fig-expt}). 
As shown in the figure, the Skyrmion lattice was stabilized at the boundary between the paramagnetic phase and a long-range
helimagnetic phase, under the application of a small external magnetic field. 
The DM interaction induced by the symmetry-breaking distortions in the B20 phase \cite{Shanavas} plays a crucial role in stabilizing the Skyrmion state in MnSi. 
Since then, other experiments have confirmed the existence of the Skyrmion state
 in a large number of chiral magnets, such as MnGe, Fe$_{1-x}$Co$_x$Si, etc \cite{Yu}. 
 More recently, experiments have shown the emergence of the
Skyrmion state in oxide heterostructures \cite{Ohuchi, Wang}, when subjected to an external electric field. 
Such magnetic Skyrmion states are shown to be tunable by the external electric field, 
which controls the DM interaction at the interface \cite{Wang}. 

In solids, the Skyrmions often form a periodic lattice (the Skyrmion crystal), rather than occurring as a single isolated Skyrmion.  The Skyrmion crystal has a lattice structure, different from the atomic lattice of the host with a lattice spacing $\sim 50 -100$ times larger than that of the underlying atomic lattice.
%A number of cases has been discovered recently. 
%
The Skyrmion crystal phase can be detected by neutron scattering and  usually small-angle neutron scattering (SANS)  is employed for this purpose.
%. Since the length scale of the SkX ($\sim$ nm) is larger than the lattice spacing ($\sim$ \AA),  small-angle neutron scattering (SANS) measurements are usually employed. 
An alternative technique is the resonant X-ray scattering.
While these are momentum-space techniques, other methods such as the
scanning probe microscopy are suitable for the direct  real-space detection of the Skyrmion crystal or the isolated Skyrmions at the nanometre scale.

%:Table 1
% \begin{table} [t]    
%\caption{List of transition temperatures (T$_N$) and helical periods ($\lambda$) of helimagnets.}
%\centering
%\setlength{\tabcolsep}{12pt}
% \begin{tabular}{c c c}
%\hline
%Materials & T$_N$ (K) & $\lambda$ (nm)\\
%            & ($\frac{\hbar}{e} \Omega^{-1}$cm$^{-1}$) \\
%\hline\hline
%MnSi (Bulk) & 30 & 18 \\
%MnSi (Thin film) & 45 & 8.5 \\
%MnGe (T = 20 K) & 170 & 3 \\
%Mn$_{0.5}$Fe$_{0.5}$Ge & 185 & 14.5 \\
%\hline
%\end{tabular} 
%\label{tab2} 
%\end{table}

\section{Skyrmion Formation: The Dzyaloshinskii-Moriya mechanism}

In this Section, we describe the formation of the Skyrmion state due to the DM interaction, which is a key mechanism. 
For the Skyrmion state to form, its energy must be lower than other competing magnetic states. For this to happen, certain types of magnetic interactions need to be present. 
According to the scaling argument of Derrick \cite{Derrick}, formation of the Skyrmion state  requires an odd power of the spatial gradient of $\vec n(\vec r)$ (note that $\vec n$ is simply the direction of the spin $\vec S$) in the energy expression. 
In these models, the magnitude of $\vec S$ remains the same everywhere in space, but their orientation can change depending on the interaction.  

The simplest model that can host the Skyrmion state has the DM interaction, which contains the gradient
$\vec \nabla \vec n$ to the first power.
Using familiar notations, the Hamiltonian reads
\begin{eqnarray}\label{SH}
 {\mathcal H} & = & J \sum_{ij} \vec{S}_i\cdot \vec{S}_j + \sum_{ij} \vec D_{ij} \cdot (\vec{S}_i \times \vec{S}_j)
 - A_s\sum_i(S_i^z)^2 - B \sum_i S_i^z,
\end{eqnarray}
where we have considered a spin model for a 2D system, where $J$ is the FM Heisenberg exchange interaction, $D$ is the DM interaction, $A_s$ is the single-ion anisotropy, and $B$ is an external magnetic field. It can be immediately seen by inspection of the 
Hamiltonian  in the continuum limit, Eq. (\ref{em}),  that the DM interaction contains the only odd-power gradient term
and therefore without it, a Skyrmion can not exist.

%:Fig 4
\begin{figure}[tb]
 \begin{center}
 \includegraphics[scale=.45]{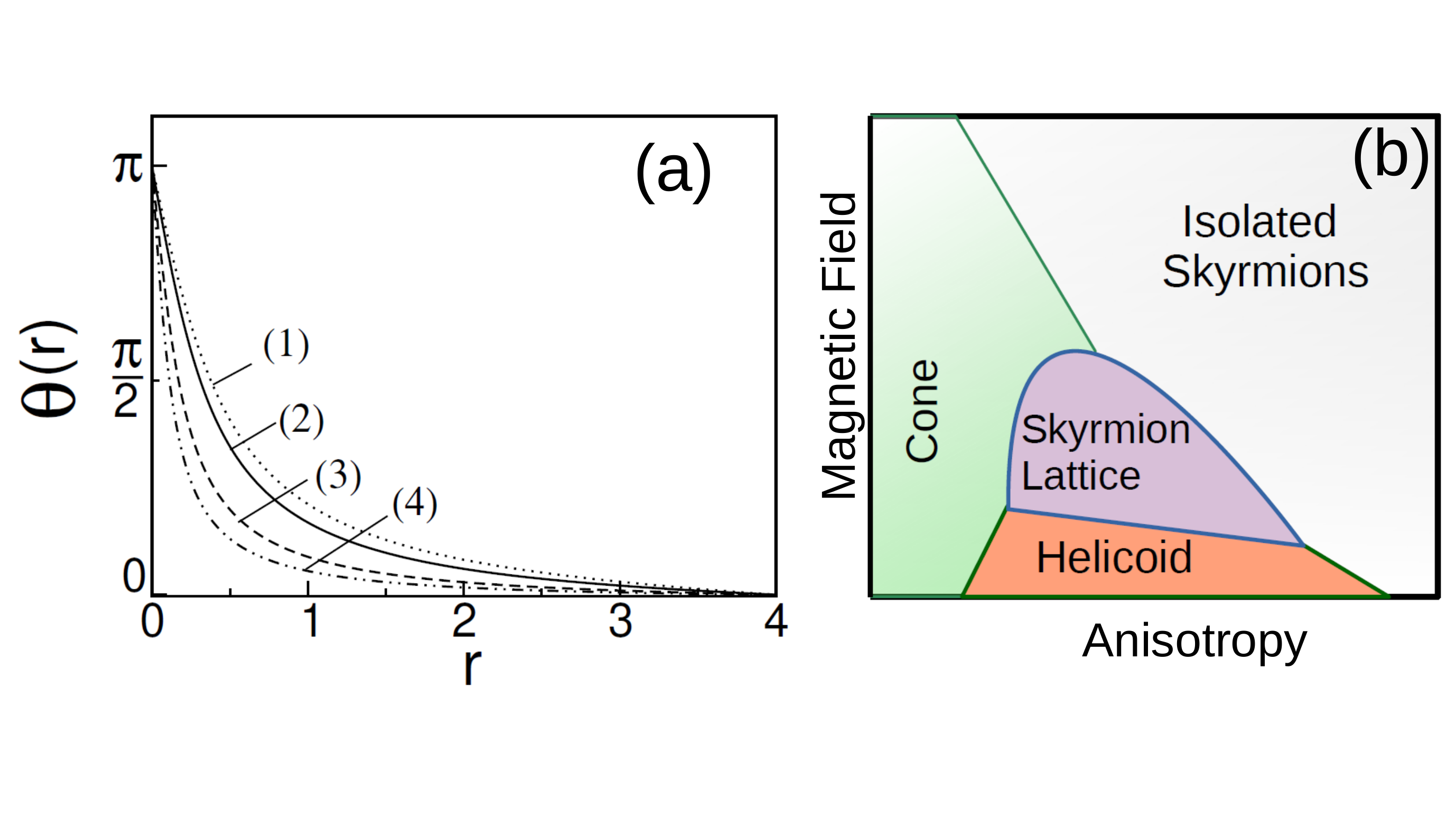}
\caption{(a) Skyrmion solution of the second order differential equation (\ref{em1})  for the parameters ($\alpha, \beta, \gamma$): (1) (0.2, 0.0, 0.1), (2) (0.2, 0.1, 0.0), (3) (0.2, 0.1, 0.1), and (4) (0.2, 0.1, 0.2).  
(b) The schematic phase diagram, adopted from Ref. \cite{Bogdanov},  showing the equilibrium states for the model spin Hamiltonian (\ref{SH}) of a cubic helimagnet as a function of applied magnetic field (B) and uniaxial anisotropy ($A_s$).
% $\alpha and \beta (\gamma$) are in arbitrary units of ($1/{\rm length}$) and ($1/{\rm length}^2$) respectively.
}
\end{center}
\label{fig4}
\end{figure}

For the purpose of illustration, we take the DM interaction of the form\cite{Mohammad}
$\vec D_{ij} = D \  \hat r_{ij} \times \hat z$, which may arise from the   Rashba SOC
at a 2D interface \cite{Mohammad}, and write the Hamiltonian (\ref{SH})
in the continuum limit. The result is
\begin{eqnarray}\label{em} \nonumber 
 E&=& \int d^2r \Big [\frac{J}{2}   \sum_{\mu = x,y,z} (\vec \nabla n^\mu)^2 +  D (n^z \partial_x n^x- n^x \partial_x n^z -n^y \partial_y n^z + n^z \partial_y n^y)\\ 
 && -A_s (n^z)^2 - Bn^z\Big ],
\end{eqnarray}
where $\vec \nabla \equiv \hat x \partial_x + \hat y \partial_y$. Note that if $D = 0$, then a uniform ferromagnetic state ($\vec n $ along $\hat z$) has the lowest energy,
since any spatial deviation of the spin direction from $\hat z$  increases the total energy, and therefore any other state including the Skyrmion state would have a higher energy. 
This is a simple consequence of Derrick's stability criterion, since if $D = 0$ in the energy expression (\ref{em}), then no odd power of $\vec \nabla \vec n$ is present anymore.
Substituting the spin texture for the Skyrmion, Eq. (\ref{n}), in the energy expression (\ref{em})  and after some straightforward algebra, we get
the expression for the energy of the Skyrmion state
\begin{eqnarray}\label{em1}
 E &= & 2\pi \int r dr \Big[ \frac{J}{2} (\dot \theta_r^2 +\sin^2\theta/r^2) + D (\dot \theta_r + \sin 2\theta/2r) - A_s \cos^2\theta -B \cos\theta\Big],
\end{eqnarray}
where we have considered a Skyrmion with unit vorticity ($m = 1$).

The problem now boils down to the minimization of the energy, which can be done using the standard
methods of the calculus of variations, which leads to the  Euler equation 
\begin{eqnarray}\label{EL}
 \frac{d} {dr} \Big (\frac{\partial F}{\partial \dot \theta_r} \Big) = \frac{\partial F}{\partial \theta}, 
\end{eqnarray}
where  $F \equiv F(\theta,\dot \theta_r,r)$ is the integrand in  (\ref {em1}). 
Taking the derivatives, we immediately get a second-order differential equation
\begin{eqnarray}\label{em2}
 \ddot \theta_r + \dot \theta_r/r + \alpha \sin^2\theta/r -\beta \sin 2\theta -\sin 2\theta/r^2 - \gamma \sin \theta= 0,
\end{eqnarray}
where we have defined the scaled  parameters $\alpha = 2D/J$, $\beta = A_s/J$, and $\gamma = B/J$. 
These are the scaled DM, anisotropy, and the magnetic field parameters, respectively. 

The solution of the differential equation is obtained with the finite difference method with the boundary conditions: $\theta (r = 0) = \pi$ and $\theta (r \rightarrow \infty) = 0$, with various values of the parameters $\alpha$, $\beta$ and $\gamma$. The results are shown in Fig. \ref{fig4} (a), which indicate the formation of an isolated Skyrmion state. 
Note that the magnetic field (parameter $\gamma$) aids in the formation of the Skyrmion state, and as the strength of the field is increased, the size of the Skyrmion becomes smaller and smaller.

So far we have discussed the role of the DM interaction in the formation of an isolated Skyrmion state in 2D. In reality, other spin textures such as the cone phase and the helicoid structure can also be stabilized. The stabilizations of these different magnetic states as a function of external magnetic field and anisotropy are   discussed in Ref. \cite{Bogdanov} in the context of cubic helimagnets. A schematic of this phase diagram is shown in Fig. \ref{fig4} (b).

%\section{Skyrmion Formation: The Dipolar interaction}

\section{Dynamics of Electrons in the presence of a Skyrmion: The topological magnetic field}
\label{sec6}

In this Section, we discuss the forces acting on an itinerant electron as it moves through the Skyrmion state, with the local space fixed magnetic moments
defined by  $\vec n (\vec r)$. 
The topological electromagnetic fields are effective fields that the electron sees on account of its motion in the network of the localized moments.
We assume a strong Hund's  coupling, which means that the only possible spin state of the electron is the one parallel to the magnetization vector of the local spin. 
Thus as the electron moves about in space, its spin must conform to the local magnetization direction. An effective electromagnetic field arises, even though no external electric or magnetic field has been applied. 
These are the so called ``emergent" or ``topological" fields that the electron experiences because the local magnetization vector changes from point to point. 

Let us consider an electron moving in a spatially varying spin texture of the Skyrmion state, with the local magnetization 
vector $ \vec n(\vec r) $ 
 described by ($\theta, \phi$) in the spherical coordinates (see Fig. \ref{polar}).
We take  the large Hund's coupling limit ($J_H \rightarrow \infty$, typically a few eV in the solid),
so that the  electron spin is everywhere parallel to the local moment as it moves about in the presence of the Skyrmion. 
Thus the spin wave function of the electron can be written as
\begin{eqnarray}     \label{chi}  
 |\chi (\vec r) \rangle &= 
\left[
{\begin{array}{*{20}c}
       \cos \frac{\theta(r)}{2} e^{-i\phi(\alpha)}  \\
     \sin\frac{\theta(r)}{2}  \\
\end{array} }  \right]
\end{eqnarray} 
in the global spin up ($ \uparrow $) and down ($\downarrow$) basis set.
As a result, the spin of the electron rotates spatially as if it experiences an effective magnetic field $\vec B (\vec r)$ 
and precesses about this magnetic field.
This physical picture is easy to follow from the kinetic energy of the conduction electron moving in a Skyrmion spin texture $\vec n(\vec r)$,
\begin{equation}
 E_{kin} = \langle \Psi | (-\hbar ^2 \nabla ^2/2m)| \Psi \rangle,
\end{equation}
where the wave function is a direct product of a spatial part and a spin part, viz.,
\begin{equation}
|\Psi\rangle = \psi (\vec r) \chi (\vec r \sigma).
\end{equation}
Writing $\nabla^2 = \partial^2_r + \frac{1}{r} \partial_r + \frac{1}{r^2} \partial^2_\alpha$ and
 $|\Psi \rangle = \psi (r,\alpha) (\cos \frac{\theta(r)}{2} e^{-i\phi(\alpha)} |\uparrow \rangle + \sin\frac{\theta(r)}{2} |\downarrow \rangle)$,
 and taking into account 
  the symmetry of the Skyrmion 
  i. e., $\theta \equiv \theta(r)$ and $\phi \equiv \phi(\alpha)$, after some algebra we get the kinetic energy of the electron 
\begin{eqnarray}\nonumber
 E_{kin} & = & \frac{-\hbar ^2}{2m} \int d^2r \Big[\psi^\star \nabla ^2 \psi - |\psi|^2 \Big ( \frac{\dot \theta_r^2}{2} 
 + \frac{\dot\phi^2_\alpha \cos^2\theta/2}{r^2} \Big)-i\frac{2 \dot\phi_\alpha \cos^2\theta/2}{r^2} \psi^\star \dot \psi_\alpha \Big] \\ 
 & = & \frac{1}{2m} \int d^2 r~ \psi^\star  \Big[ (\vec p - e \vec A/c)^2  + V \Big] \psi ,
%  & \equiv & \frac{1}{2m} \int d^2 r~ \psi^\star  \Big[ (\vec p - e \vec A^\prime/c)^2   \Big] \psi, 
 \label{ke}
\end{eqnarray}
%
%:Fig 5
\begin{figure}[tb]
 \begin{center}
\includegraphics[scale=.3]{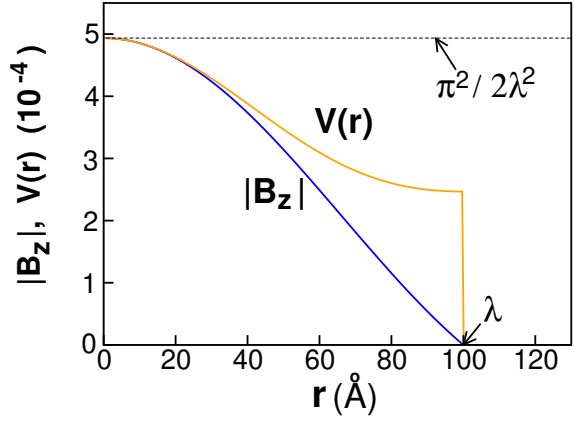}
\caption{Spatial variations of the topological magnetic field (\ref{bz}) and the scalar potential $V(r)$ (\ref{vr}) for the Skyrmion profile, Eq. (\ref{sk}). The radius of the Skyrmion $\lambda$ is taken to be 100 \AA. For this figure, we have taken $\hbar = c = e =1$.}
\end{center}
\label{fig5}
\end{figure}
where $\vec p \equiv -i \hbar \vec \nabla$, $ e < 0$ is the charge of the electron, and the derivatives are indicated by the subscripts, e.g.,  
$\dot \psi_\alpha \equiv \partial \psi /\partial \alpha$.
Also, here, $V =  4^{-1} \hbar^2 \Big ( r^{-2} \dot \phi_\alpha^2   \sin^2\theta  + \dot \theta_r^2 \Big)$,
$\vec A = \frac{\Phi_0}{2\pi r}  (\dot \phi_\alpha \cos^2\theta/2 ) (\hat r \times \hat z)$ with $V$ and $\vec A$ being the scalar and the vector potentials respectively,
and  $\Phi_0 = hc/e$ is the flux quantum  in cgs units.  
The scalar potential corresponds to a radially outwards force, while the vector potential $\vec A$ leads to the 
topological magnetic field 
\begin{eqnarray}\label{bz1}
 \vec B (\vec r)  & = & \vec \nabla \times \vec A  
            = \hat z  (\partial_x A_y - \partial_y A_x) 
                 = \hat z \frac{\Phi_0}{2\pi} \frac{\dot \theta_r \dot \phi_\alpha \sin \theta}{2r} \\ \nonumber
& = & \frac{\Phi_0}{4\pi} \vec n \cdot \Big ( \partial_x \vec n \times \partial_y \vec n \Big) \hat z.
 \hspace{5mm} {\rm (topological \ magnetic \ field)}
\end{eqnarray}

% \begin{equation}
% \vec B (\vec r)  = \vec \nabla \times \vec A 
%                & = & \hat z  (\partial_x A_y - \partial_y A_x) \\ \nonumber
%                 & = & \hat z \dot \theta_r \dot \phi_\alpha \sin \theta/2r \\
% = \frac{1}{2} \vec n \cdot \Big ( \partial_x \vec n \times \partial_y \vec n \Big) \hat z.
% \hspace{5mm} {\rm (topological \ magnetic \ field)}
%\end{equation}
%
%
We pause here to consider the special case where the magnetization vector $\vec n$ points along 
$\hat z$ everywhere ($\theta = \phi = 0$), for example, for the case of the ferromagnetic state.
Then clearly both the scalar and vector potentials are zero from the above expressions and the electron moves as a free particle. 
In the Skyrmion state, however, the electron experiences a spatially varying topological magnetic field.
 Let us assume the Skyrmion profile 
 \begin{eqnarray}\label{sk}\nonumber
  \theta(r) & = &\pi (1-r/\lambda),~ r \le \lambda \\ \nonumber
            & = & 0,~  r > \lambda\\ 
  \text {and ~~~~~} \phi (\alpha) &=& \alpha, 
\end{eqnarray}
where $\lambda$ denotes the Skyrmion radius.
Using these functional forms of $\theta(r)$ and $\phi (\alpha)$, 
it is easily seen that the topological  magnetic field is along $\hat z$ with the magnitude
 \begin{eqnarray}\label{bz}             \nonumber
  B_z & = & -\frac{\Phi_0 }{4 \lambda r} \sin (\pi r/\lambda), \ {\rm if} \ \  r \le \lambda \\ \nonumber
            & = & 0 ~,   \ {\rm if} \ \ r > \lambda, \\  
 \end{eqnarray}
and
 \begin{eqnarray}\label{vr}       \nonumber
  V(r) & = & 4^{-1}\hbar^2\Big[ \frac{\sin^2 (\pi r/\lambda)}{r^2} + \frac{\pi^2}{\lambda^2} \Big], \ {\rm if} \ \  r \le \lambda \\ \nonumber
            & = & 0 ~,   \ {\rm if} \ \ r > \lambda. \\  
 \end{eqnarray}
 The spatial variation of these quantities are shown in Fig. \ref{fig5}. As can be seen from these plots, at the centre of the Skyrmion ($r \rightarrow 0$) $|B_z| = V(r) = \pi^2/2\lambda^2$. For typical Skyrmion radius $\lambda = 100 $      
 \AA, this leads to a large topological magnetic field ($\sim 10^2$  Tesla), which thus provides a unique platform to study the high
magnetic field response of electrons \cite{Bruno}.  

{\it Tight-Binding description} -- In the above analysis, we obtained the topological magnetic field from a continuum description, while 
in the solids, the Skyrmions exist of course on crystal lattices. 
A different way of obtaining the topological magnetic field is to consider the phase accumulated by the traveling electron in the lattice. 
As the electron travels, its spin must follow the fixed local magnetization direction $\vec n (\vec r)$ adiabatically, as we already discussed.
The hopping matrix element between two lattice sites $a$ and $b$, in the tight-binding theory becomes
\begin{eqnarray}
t_{ab} = \langle \Psi_a | H | \Psi_b \rangle = \langle \psi_a | H | \psi_b \rangle \times \langle \chi_a | \chi_b \rangle = t e^{i \eta},
\label{tab}
 \end{eqnarray}
where $t = \langle \psi_a | H | \psi_b \rangle$ is the electron hopping without the spin texture 
and in the last equality, we have assumed a slowly varying spin texture 
 (i. e., change in $\theta $ and $\phi $ between nearest neighbors in the lattice is assumed to be small). 
 As is well known from the seminal book by Feynman    
  \cite{Feynman}, when an electron moves in a magnetic field,
 it acquires a phase factor  (see Fig. 6), viz.,
  \begin{eqnarray} 
  t_{A \ne 0} = t_{A = 0} \times  \exp \big(\frac{i e}{\hbar c} \int_a^b \vec A \cdot \vec dr \big).
  \label{Feynman} 
  \end{eqnarray}
  Comparing Eqs. (\ref{tab}) and (\ref{Feynman}), it is clear that even though no external magnetic field is present, the adiabatic
  motion of the electron in the spin texture is equivalent to a magnetic field, the ``topological" magnetic field.
  Expressing the phase factor in Eq. ({\ref{tab}) in terms of the Skyrmion texture $\theta (\vec r)$ and $\phi (\vec r)$
  and after some straightforward algebra, one again gets the expression Eq. (\ref{bz1}) for the
  topological magnetic field $\vec B (\vec r)$.
  
%: Fig 6
\begin{figure}[tb]
\label{fig6}
 \begin{center}
\includegraphics[scale=.15]{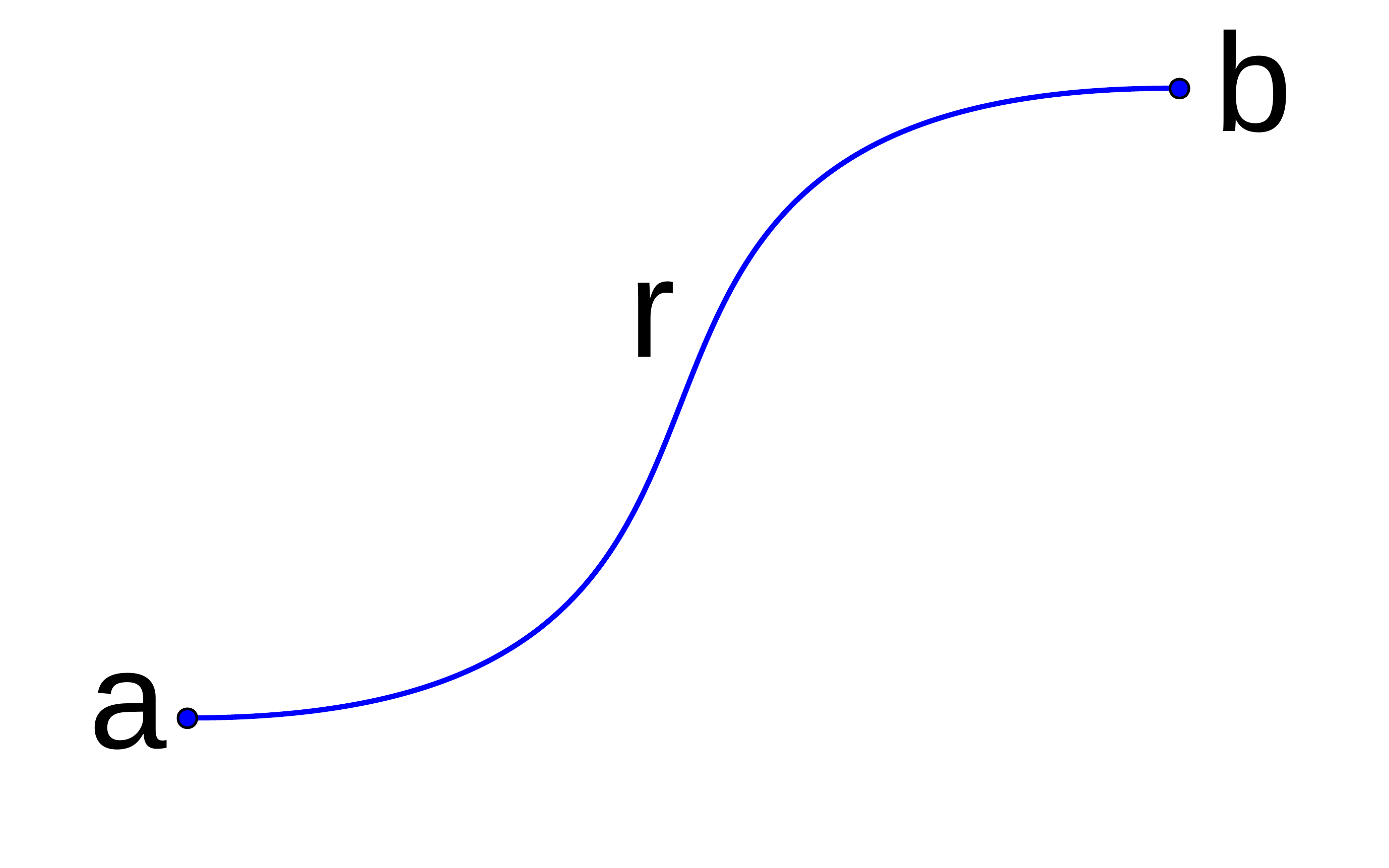}
\caption{In the presence of a magnetic field, the  amplitude to go from point $a$ to point $b$ acquires an extra phase factor 
$  \big[   \exp \big(\frac{i e}{\hbar c} \int_a^b \vec A \cdot \vec dr \big)   \big]$,
where $\vec A$ is the magnetic vector potential,  as discussed in the Feynman Lecture Notes \cite{Feynman}. 
This modifies the hopping amplitude in the  tight-binding theory as indicated in Eq. \ref{Feynman}.
}
\end{center}
\end{figure}
 
%\section{Topological Properties}  

{\it Magnetic flux through the Skyrmion} -- It is easy to show that the flux through the Skyrmion due to the topological magnetic field is an integer multiple
(the Skyrmion number $N_{sk}$) of
the flux quantum $\Phi_0$. 
Computing the flux through the 2D surface containing the Skyrmion with the magnetic field Eq. (\ref{bz1}), we have
\begin{eqnarray}
 \Phi  = \int_S \vec B \cdot d\vec S 
       =  \frac{\Phi_0}{4\pi} \int  \Big [\vec n \cdot \Big ( \partial_x \vec n \times \partial_y \vec n \Big) \Big] \ dS =
      \Phi_0 N_{sk},
\end{eqnarray}
where the Skyrmion number $N_{sk}$
%$N_{sk} = \frac{1}{4\pi} \int d\vec r \Big [\vec n \cdot \Big ( \partial_x \vec n \times \partial_y \vec n \Big) \Big]$ 
characterizes the swirling structure of the Skyrmion. 
The fact that the Skyrmion number is an integer is a simple consequence of the boundary conditions imposed on $\vec n (\vec r)$
to define the Skyrmion state.
Using the Skyrmion spin texture Eq. (\ref{n}), we can immediately see that the Skyrmion number is simply the winding number $m$
times a sign $\eta = \pm 1$,
\begin{eqnarray}\label{nsk}
 N_{sk} = \frac{1}{4\pi} \int_0^\infty dr \sin \theta(r) \dot \theta_r  \times \int_0^{2\pi} d\alpha~\dot \phi_\alpha 
= \eta m.
\end{eqnarray}
The quantity $\eta = \frac{1}{2} \cos \theta(r)|_{r = 0}^{r = \infty} = \pm 1$  is related to the magnetization direction at the origin, i. e., 
if the spins  point up  at the origin $(r = 0)$ and down at $r \rightarrow \infty$, then $\eta = + 1$, and it is $-1$ in the opposite case.

\section {Skyrmionic Polaron}

In this Section, we discuss the idea whether an electron can nucleate a Skyrmion state.
The term ``Skyrmionic polaron" has been recently coined \cite{Brey} to describe such a system, which is analogous to the standard polaron state,
which is a combination of an electron and the distorted lattice, which together form a localized state due to the electron-phonon interaction. 

Interfaces are of interest because they are fertile grounds for the experimental observation of the Skyrmions. 
It has been recently suggested that the Skyrmionic polaron may exist at the interface between two solids, aided by the so called
Rashba spin-orbit coupling \cite{Rashba}.
The idea is that
due to the broken mirror symmetry, an electric field perpendicular to the interface plane can exist, and if a large spin-orbit coupling is present, they both together
lead to the momentum-dependent
spin splitting of the electron states at a surface or interface, known as Rashba effect \cite{Rashba}.  

{\it Rashba spin-orbit interaction} --
The Rashba interaction is a relativistic effect, originating from the coupling between the spin and the orbital degrees of freedom, when
an electric field is present along a certain direction. 
This happens in solids when the mirror symmetry is broken, e.g., at the surface or at  the interface between two materials. 
The Rashba interaction is  described by  the Hamiltonian
\begin{equation}\label{Rashba}
  {\cal H}_R = \alpha_R (\sigma_x k_y - \sigma_y k_x),
\end{equation} 
where $\vec k$ is the electron momentum, $ \vec \sigma = (\sigma_x, \sigma_y, \sigma_z)$  are the Pauli matrices, and $\alpha_R$ is the strength of the Rashba interaction.
Diagonalization of (\ref{Rashba}) leads to the additional linear splitting term on top of the quadratic band structure energy,
with the result 
$\varepsilon_k = (\hbar^2k^2/2m) \pm \alpha_R k$. 
 
 The microscopic origin of the Rashba interaction is similar to the relativistic effect that leads to
 the well-known $\lambda \vec L \cdot \vec S$ term in the atoms. 
 Consider an electron moving in an electric field $E \hat z$, relevant for the Rashba problem. 
 Due to the relativistic effect, in its rest frame, the electron sees the electric field as a magnetic field
 $\vec B = -( \vec v \times \vec E) / c^2$,
 which couples with the spin magnetic moment.
 The interaction energy is given by 
 $ {\cal H}_R = -\vec M \cdot \vec B  =  (g \mu_B \vec \sigma / 2) \cdot   (\vec v \times \vec E) / c^2 $.
 Writing $\vec v = \hbar \vec k / m$ and rearranging the terms, one immediately finds the Rashba Hamiltonian (\ref{Rashba}), with the Rashba coefficient $\alpha_R$ is expressed in terms of the 
 various fundamental quantities.
 In addition, one can easily show that if instead of the uniform field along a fixed direction, the nuclear electric field along the radial direction is used, we
 would obtain the well-known $\lambda \vec L \cdot \vec S$ term in atomic physics.

The Rashba interaction leads to many interesting phenomena in condensed matter physics, but relevant for the present case is the fact that
it produces a Dzyaloshiski-Moriya interaction $\vec D \cdot (\vec S_i \times \vec S_j)$ between two local spins \cite{Mohammad}, which facilitates the formation of a Skyrmion state by supporting a non-collinear alignment of the spins. 
So, the question that naturally arises is whether a single electron with the Rashba interaction in a 2D system can nucleate a Skyrmion state for the local spins. 

{\it Formation of the Skyrmionic polaron--}
Consider the following continuum model for a single electron moving in the Skyrmionic spin texture $\vec n (\vec r)$ in the presence
of the Rashba interaction. The Hamiltonian is
\begin{equation} \label{H2}
{\cal H} = \frac{\hbar^2 k^2} {2m}    - J \vec n \cdot \vec \sigma    + (\sigma_x k_y - \sigma_y k_x)    - b_z \int n_z (\vec r) \ d^2r,
\end{equation}
where we have added a magnetic field $b_z$ normal to the plane for generality because in certain situations, the Skyrmion state is stabilized by a magnetic field,
and the exchange coupling $J$ between the local spin texture and the electron spin will be taken as $\infty$.
% Here,   $\alpha_R$ is the strength of the Rashba interaction, $ \vec \sigma = (\sigma_x, \sigma_y, \sigma_z)$  are the Pauli matrices, $\vec k$ is the electron momentum, and the exchange coupling between the local spin texture and the electron spin $J$ with be taken as $\infty$.
The large $J \rightarrow \infty$ condition effectively renders this to a spinless problem as the electron spin is always parallel to the spin texture $\vec n (\vec r)$ everywhere as the electron moves around. Switching to the spin basis to be along the local moments (Eq. \ref{chi}),
and modeling the Skyrmion texture $\vec n (\vec r)$  again by the form:
$ \theta(r)  = \pi (1-r/\lambda),~ r \le \lambda$, and $0$ otherwise, and $\phi (\alpha) = \alpha$,
the spinless electron experiences the magnetic field as well as a static potential given by Eqs. (\ref{bz}) and (\ref{vr}), respectively.

We can then proceed to obtain the ground state configuration of the Skyrmion by minimizing the total energy of the  Hamiltonian (\ref{H2}) 
from the variational method, treating the Skyrmion radius $\lambda$ as the single variational parameter.
The average topological magnetic field has the value
$\bar B = \Phi_0/(\pi \lambda^2)$ and the corresponding magnetic length $l = \lambda / \sqrt 2$. 
The quantized electron states of the electron moving in a magnetic field are the Landau levels, with the lowest energy being $ \hbar \omega_c / 2= \hbar^2/(m \lambda^2)$, $\omega_c$ being the cyclotron radius. 
The potential energy may be approximated as $\langle V(r) \rangle = V (l) $, where we simply evaluate the expression (\ref{vr}) for $V(r)$ at $r = l$.

The last term in the Hamiltonian (\ref{H2}) is the Rashba term. We evaluate the Rashba energy for the symmetric wave function $\psi (r)$, where there is no angular dependence $\alpha$, with the result
\begin{equation}
H_R = \langle \psi (r) \uparrow | \alpha_R (k_x \sigma_y - k_y \sigma_x)| \psi(r)\uparrow\rangle = \frac{\alpha_R}{2}  \int |\psi|^2 \times 
(\dot \theta_r - \frac{\dot \phi_\alpha}{r} \sin \theta) \ d^2r,
\end{equation}
where $\uparrow$ denotes the spin aligned along the local moment and $\vec k \equiv -i \vec \nabla $. Evaluating this integral and putting together all energy terms, we finally arrive at the result
\begin{equation} \label{Elambda}
E(\lambda) = \frac{\hbar^2 a}{2m \lambda^2} + 
 \pi b_z \lambda^2 (1-\frac{4}{\pi^2}) -\frac{\alpha_R b}{\lambda},
\end{equation}
where $a = \pi^2/4 + 2 +  2^{-1}\sin^2 (\pi/\sqrt 2) $ and $b = \pi / 2 + 2^{-1/2} \sin (\pi/ \sqrt 2)$ are numerical constants, and the
three terms are respectively the kinetic energy, the external magnetic field energy, and the Rashba energy. 
Minimizing the energy $d E(\lambda)/ d \lambda = 0 $ with no external magnetic field ($b_z = 0$), we obtain the Skyrmion radius and the corresponding ground state energy. 
The result is
\begin{equation}
\lambda_0 \approx \frac {2\hbar^2}{m\alpha_R}, \ \ \ \ \ {\rm and} \ \ E_0 \approx -  \frac {3m \alpha_R^2}{4\hbar^2}.
\end{equation}
Clearly, without any Rashba interaction ($\alpha_R = 0$), the Skyrmion radius $\lambda_0$ is infinity, indicating that there is no Skyrmion state, while the presence of the Rashba term favors the formation of the Skyrmion state with the binding energy $E_0$. 
The result is that the electron nucleates the Skyrmion state and in turn becomes bound inside it,
forming thereby the ``Skyrmionic polaron." From  Eq. (\ref{Elambda}), it can be seen that an external magnetic field further aids in the formation of the Skyrmionic polaron, making its radius smaller and the binding energy larger.

\section{Applications and Future Perspective}

The topic of the magnetic Skyrmions has been rapidly developing over the past few years, both from the viewpoint of fundamental physics as well as the prospect for technological applications \cite{Finocchio}.
The Skyrmion spin texture is topologically stable with small thermal and quantum fluctuations, which makes it suitable for applications in memory devices.
The unusual electron dynamics and transport properties such as the Topological Hall effect (THE) \cite{Nagaosa,Yu,Neubauer} could have important application in spintronics devices.
 While the first discovery of the Skyrmions occurred in bulk materials with chiral magnetic
interactions, the realization that they can also be stabilized at the interfaces of magnetic multilayers \cite{Ranieri} has opened up additional potential opportunities,
including novel pathways for Skyrmion generation and manipulation.

A promising application involves the so called Topological Hall effect (THE).
In the well-known classical Hall effect, discovered by Edwin H. Hall in 1879,
when a current carrying conductor is placed in a magnetic field, the charges experience a Lorentz force in a direction perpendicular to both the magnetic field and the current 
flow. In contrast to this, in the Topological Hall effect, it is the topological magnetic field of a non-collinear spin system such as the Skyrmion
gives rise to the Lorentz force on the conduction electron, resulting in a different type of Hall effect. 
At the same time, the motion of the Skyrmion itself leads to a temporal variation of the topological magnetic field. This, in turn, induces an electromotive force or potential according to the Faraday's law. This topological electric field gives an additional contribution to the Hall effect. This has been observed recently in epitaxial thin films and even in nanowires, demonstrating the physical reality of the topological electromagnetic fields in solids \cite{eemf}.  

%In the past decades, a variety of Hall effects has been discovered. In contrast to the other Hall effects (e.g., {\it Anomalous Hall effect}), which are driven by the spin-orbit scattering, very recently a new type of Hall effect has been discovered in frustrated non-collinear spin systems, related to the topology of the magnetization texture. This new type of Hall effect is known as THE, often hosted by the Skyrmions.

A large number of Skyrmion based innovative devices have been recently proposed \cite{Kang2,Liu,Zhang},
%such as Skyrmion racetrack memory [308], Skyrmion nano-oscillators [257, 311, 312], Skyrmion transistors [313],Skyrmion logics [316] etc, 
although there remains several important issues to be resolved \cite{Finocchio} before  practical applications can be made. A notable Skyrmion based memory device is the Skyrmion racetrack memory, which has received considerable attention \cite{Kang2}. While a prototype racetrack memory has been successfully demonstrated, a practical electrical read-out scheme still remains to be developed. 
Another intriguing idea is the use of a single Skyrmion as the information bit \cite{Hagemeister}.
Research on the magnetic Skyrmions is currently in its infancy, with rapid development in the fundamental physics and applications expected in the coming years.

%To conclude, from the application perspective, the field is still young. In this context, the nanoscale fabrication of samples, design and demonstration of Skyrmion logic circuits are definitely some important steps towards the  future realization of functional devices, enabled by fundamental understanding of the basic physics of the Skyrmions.  

\section{Acknowledgments}
We thank the U.S. Department of Energy, Office of Basic Energy Sciences, Division of Materials Sciences and Engineering for financial support under Grant No. DEFG02-00ER45818.

%========================================================
%\newpage


\begin{thebibliography}{99}

\bibitem{Skyrme} Skyrme, T. H. R. 1962, ``A unified field theory of mesons and baryons", Nucl. Phys. {\bf 31}, 556-569

\bibitem{Bogdanov1989} Bogdanov, A. N. and Yablonskii, D. A. 1989, ``Thermodynamically stable vortices in magnetically ordered crystals: 
The mixed state of magnets", Soviet Phys. JETP {\bf 68}, 101-103


\bibitem{Nagaosa} For a review of recent works, see: Nagaosa, N. and Tokura, Y. 2013, ``Topological properties and dynamics of magnetic Skyrmions",  Nature Nanotechnol. {\bf 8}, 899-911


\bibitem{Jacob} Mechelen, T. V.  and  Jacob, Z. 2019, ``Viscous Maxwell-Chern-Simons theory for topological electromagnetic phases of matter", arXiv:1910.14288v1 (2019); Mechelen, T. V.  and  Jacob, Z. 2019, ``Nonlocal topological electromagnetic phases of matter", Phys. Rev. B {\bf 99}, 205146. 


%-----------





%---------------

\bibitem{Garel} Garel, T. and Doniach, S. 1982, ``Phase transitions with spontaneous modulation - the dipolar Ising ferromagnet," Phys. Rev. B {\bf 26}, 325-329

\bibitem{Heinze} Heinze, S. {\it et al.} 2011, ``Spontaneous atomic-scale magnetic Skyrmion lattice in two dimensions," Nature Phys. {\bf 7}, 713-718

\bibitem{Okubo} Okubo, T., Chung, S. and Kawamura, H. 2012, ``Multiple-q states and the Skyrmion lattice of the triangular-lattice heisenberg antiferromagnet under magnetic fields," Phys. Rev. Lett. {\bf 108}, 017206 

%\bibitem{Yu2010} Yu, X. Z. , Onose, Y.,  Kanazawa, N.,  Park, J. H.,  Han, J. H., Matsui, Y.,  Nagaosa, N. and  Tokura, Y.  2010, ``Real-space observation of a two-dimensional skyrmion crystal", Nature {\bf 465}, 901-904.

\bibitem{Brey} Brey, L. 2017,`` Magnetic Skyrmionic Polarons," Nano Lett. {\bf 17}, 7358-7363

\bibitem{MnSi} M\"uhlbauer, S., Binz, B. , Jonietz, F. ,  Pfleiderer, C. , Rosch, A., Neubauer, A., Georgii, R.  and Boni, P. 2009, `` Skyrmion lattice in a chiral
magnet," Science {\bf 323}, 915 

\bibitem{Yu2010} Yu, X. Z. , Onose, Y.,  Kanazawa, N.,  Park, J. H.,  Han, J. H., Matsui, Y.,  Nagaosa, N. and  Tokura, Y.  2010, ``Real-space observation of a two-dimensional skyrmion crystal", Nature {\bf 465}, 901-904



%\bibitem{MnSi} M\"uhlbauer, S., Binz, B. , Jonietz, F. ,  Pfleiderer, C. , Rosch, A., Neubauer, A., Georgii, R.  and Boni, P. 2009, `` Skyrmion lattice in a chiral magnet," Science {\bf 323}, 915 

\bibitem{Shanavas} Shanavas, K. V. and Satpathy, S. 2016, ``Electronic structure and the origin of the Dzyaloshinskii-Moriya interaction in MnSi," Phys. Rev. B {\bf 93}, 195101


\bibitem{Yu} Yu, X. Z. ,  Kanazawa, N.,  Onose, Y.,  Kimoto, K.,  Zhang, W. Z.,  Ishiwata, S.,  Matsui, Y. and  Tokura, Y. 2011,  ``Near room-temperature formation of a Skyrmion crystal in thin-films of the helimagnet FeGe," Nat. Mater. {\bf 10}, 106
%\bibitem{Yu} Yu, X. Z. {\it et al.} 2012, ``Skyrmion flow near room temperature in an ultralow current density," Nature Commun. {\bf 3}, 988

\bibitem{Ohuchi} Ohuchi, Y. {\it  et al.}  2018, ``Electric-field control of anomalous and topological Hall effects in oxide bilayer thin films,"  Nat. Commun. {\bf 9}, 213



\bibitem{Wang} Wang, L. {\it  et al.} 2018, ``Ferroelectrically tunable magnetic Skyrmions in ultrathin oxide heterostructures", Nat. Mater. {\bf 17}, 1087-1094



%\bibitem{Blugel} Heinze, S., von Bergmann, K., Menzel, M., Brede, J., Kubetzka, A., Wiesendanger, R., Bihlmayer, G. and Bl\"ugel, S. 2011, `` Spontaneous atomic-scale magnetic Skyrmionlattice in two dimensions," Nat Phys, {\bf 7}, 713-718

%\bibitem{Blugel2017} Hoffmann, M., Zimmermann, B., M\"uller, G. P.,  Sch\"urhoff, D., Kiselev, N. S., Melcher , C. and Bl\"ugel, S. 2017,``AntiSkyrmions stabilized at interfaces byanisotropic Dzyaloshinskii-Moriya interactions," Nat Commun, {\bf 8}, 308

\bibitem{Derrick} Derrick, G. H. 1964, ``Comments on Nonlinear Wave Equations as Models for Elementary Particles", J.Math. Phys., {\bf 5}, 1252

\bibitem{Mohammad} Mohammad, M. V. and Satpathy, S. 2018, ``Dzyaloshinskii-Moriya interaction in the presence of Rashba and Dresselhaus spin-orbit coupling", Phys. Rev. B, {\bf 97}, 094419 


\bibitem{Bogdanov} Wilson, M. N. , Butenko, A. B.,  Bogdanov, A. N. and Monchesky, T. L. 2014, ``Chiral Skyrmions in cubic helimagnet films: The role of uniaxial anisotropy", Phys. Rev. B, {\bf 89}, 094411 

\bibitem{Bruno} Bruno, P., Dugaev, V. K. and Taillefumier, M. 2004, ``Topological Hall effect and Berry phase in magnetic
nanostructures", Phys. Rev. Lett., {\bf 93},  096806

\bibitem{Feynman} The Feynman Lectures on Physics Vol III Ch. 21, ``The Schr\"odinger Equation in a Classical Context: A Seminar on Superconductivity".


\bibitem {Rashba}  Rashba, E. I. 1960,  `` Properties of semiconductors with an extremum loop. 1. Cyclotron
and combinational resonance in a magnetic field perpendicular to the plane of
the loop", Sov. Phys. Solid State {\bf 2}, 1109 ; 
Bychkov, Y. A. and Rashba, E. I. 1984,  `` Oscillatory effects and the magnetic susceptibility of
carriers in inversion layers", J. Phys. C {\bf 17}, 6039 

%%\bibitem{Jenietz} Jonietz, F. {\it et al.} 2010,  ``Spin transfer torques in MnSi at ultralow current densities," Science {\bf 330}, 1648-1651

%\bibitem{Yu} Yu, X. Z. {\it et al.} 2012, ``Skyrmion flow near room temperature in an ultralow current density," Nature Commun. {\bf 3}, 988

\bibitem{Finocchio} Finocchio, G., B\"uttner, F. , Tomasello, R.,  Carpentieri, M., Kl\"aui, M. 2016 `` Magnetic Skyrmions: from
fundamental to applications", J Phys D Appl Phys, {\bf 49}, 423001

\bibitem{Neubauer} Neubauer, A. {\it et al.} 2009, ``Topological Hall effect in the A phase of MnSi", Phys. Rev. Lett. {\bf 102}, 186602

%%\bibitem{Yufan} Yufan Li, Y. {\it et al.} 2013,  ``Robust formation of Skyrmions and topological Hall effect anomaly in epitaxial thin films of MnSi," Phys. Rev. Lett. {\bf 110}, 117202

\bibitem{Ranieri} Ranieri, E. D. 2014, ``Skyrmions at the interface",  Nature Nanotechnology {\bf 101}, 10.1038

\bibitem{eemf} Schulz, T. {\it et al.} 2012, ``topological electrodynamics of Skyrmions in a chiral magnet", Nature Phys. {\bf 8}, 301–304

%%\bibitem{QTHE} Hamamoto, K., Ezawa, M. and Nagaosa N. 2015, `` Quantized topological Hall effect in Skyrmion crystal'', Phys. Rev. B, {\bf 92}, 115417
%%%%%%%%%%%%%%%%%%%%%%%%%%%%%%%%%%%%%%%%%%%%%%%%%%%%%%%%%%%%%%%%%%
%\bibitem{halli94} Halliday D., Resnick R., and Walker J. 1994 \textit{Fundamentals of Physics}  (New Delhi: Asian Books Pvt. Ltd.) 

%\bibitem{pati}  Supriya D Patil and S V Anekar, 2014, ``Effect of Different Parameters and Storage Conditions on Liquid Jaggery Without Adding Preservatives'', Intnl J of Res in Engg and Tech, \textbf{12}, 280 - 283

%\bibitem {Rashba}  Rashba, E. I. 1960,  `` Properties of semiconductors with an extremum loop. 1. Cyclotron and combinational resonance in a magnetic field perpendicular to the plane of the loop," Sov. Phys. Solid State {\bf 2}, 1109 ; Bychkov, Y. A. and Rashba, E. I. 1984,  `` Oscillatory effects and the magnetic susceptibility of carriers in inversion layers," J. Phys. C {\bf 17}, 6039 
%Y. A. Bychkov and E. I. Rashba, JETP Lett. {\bf39}, 78 (1984).

%%\bibitem{Kang} Kang, W., Huang,  Y. Q.,  Zhang,  X.C.,  Zhou, Y., Zhao, W.S. 2016, `` Skyrmion-Electronics: An Overview and Outlook," Proceedings of the IEEE, {\bf 104}, 2040-2061

%%\bibitem{Finocchio} Finocchio, G., B\"uttner,  F. ,  Tomasello, R.,  Carpentieri, M.,  Kl\"aui, M. 2016, ``Magnetic Skyrmions: from fundamental to applications," J Phys D Appl Phys {\bf 49}, 423001

%%\bibitem{Fert} Fert, A., Reyren, N.,  Cros, V.  2017, ``Magnetic Skyrmions: advances in physics and potential applications", Nat Rev Mater {\bf 2}, 17031.


%\bibitem{Ranieri} Ranieri, E. D. 2014, ``Skyrmions at the interface",  Nature Nanotechnology {\bf 101}, 10.1038

%%\bibitem{Chen} Chen, J. {\it et al.} 2019, ``Evidence for Magnetic Skyrmions at the Interface of Ferromagnet/Topological-Insulator Heterostructures,"  Nano Lett. {\bf 19}, 6144-6151

\bibitem{Kang2} Kang, W., Zheng, C.,  Huang, Y., Zhang, X., Zhou, Y., Lv, W., 
Zhao, W. 2016 ``Complementary Skyrmion Racetrack Memory with Voltage Manipulation", IEEE Electron
Device Letters, {\bf 37}, 924-927

\bibitem{Liu} Liu, R.H., Lim, W.L. , Urazhdin, S.  2015 ``Dynamical Skyrmion State in a Spin Current Nano-Oscillator with Perpendicular Magnetic Anisotropy", Phys. Rev. Lett. {\bf 114} , 137201

\bibitem{Zhang} Zhang, X. ,  Zhou, Y., Ezawa, M. ,  Zhao, G.P.,  Zhao, W. 2015 ``Magnetic Skyrmion transistor: Skyrmion
motion in a voltage-gated nanotrack", Sc.  Rep. {\bf 5}, 11369

%\bibitem{Finocchio} Finocchio, G., B\"uttner, F. , Tomasello, R.,  Carpentieri, M., Kl\"aui, M. 2016 `` Magnetic Skyrmions: from
fundamental to applications", J Phys D Appl Phys, {\bf 49}, 423001

%%\bibitem{Fert2} Fert, A., Cros, V. , Sampaio, J. , 2013 `` Skyrmions on the track, Nature Nanotechnology," {\bf 8}, 152-156

%%\bibitem{Romming} Romming, N. , Hanneken,  C.,  Menzel, M., Bickel, J.E., Wolter,  B. , von Bergmann, K. , Kubetzka, A., Wiesendanger, R. 2013 `` Writing and deleting single magnetic Skyrmions," Science, {\bf 341}, 636-639

\bibitem{Hagemeister} Hagemeister, J.,  Romming,  N., von Bergmann, K. , Vedmedenko, E.Y.,  Wiesendanger,  R., 2015 `` Stability of single Skyrmionic bits", Nat. Commun., {\bf 6}, 8455

\end{thebibliography}
\end{document}